\documentclass[aps,prd,twocolumn,nofootinbib,showpacs,superscriptaddress]{revtex4-1}

\usepackage{amsfonts}
\usepackage{amsmath}
\usepackage{amssymb}
\usepackage{bm}
\usepackage{dcolumn}
\usepackage[dvips]{graphicx}
\usepackage{graphics}
\usepackage[latin1]{inputenc}
\usepackage{latexsym}
\usepackage{rotating}
\usepackage[colorlinks=true]{hyperref}
\usepackage{xspace} 
\usepackage[usenames]{color}
\usepackage{mathrsfs}
\usepackage{multirow}
\usepackage{pifont}
\usepackage{enumitem}
\usepackage{cleveref}

\definecolor {darkgreen}{rgb}{0.2,0.7,0.2}
\definecolor{purple}{rgb}{0.5,0,0.5}

\newcommand{\be}{\begin{equation}}
\newcommand{\ee}{\end{equation}}
\newcommand{\bea}{\begin{eqnarray}}
\newcommand{\eea}{\end{eqnarray}}

\newcommand{\FLRW}{{\mbox{\tiny FLRW}}}
\newcommand{\Mink}{{\mbox{\tiny Mink}}}

\begin{document}

\title{Gravitational wave memory in $\Lambda$CDM cosmology}

\author{Lydia Bieri}
\email{lbieri@umich.edu}
\affiliation{Department of Mathematics, University of Michigan, Ann Arbor, MI 48109-1120, USA}

\author{David Garfinkle}
\email{garfinkl@oakland.edu}
\affiliation{Department of Physics, Oakland University, Rochester, MI 48309, USA}
\affiliation{Michigan Center for Theoretical Physics, Randall Laboratory of Physics, University of Michigan, Ann Arbor, MI 48109-1120, USA}

\author{Nicol\'as Yunes}
\email{nicolas.yunes@montana.edu}
\affiliation{eXtreme Gravity Institute, Department of Physics, Montana State University, Bozeman, MT 59717 USA}


\date{\today}

\begin{abstract}
We examine gravitational wave memory in the case where sources and detector are in a $\Lambda$CDM cosmology.  We consider the case where the universe can be highly inhomogeneous, but the gravitatational radiation is treated in the short wavelength approximation.  We find results very similar to those of gravitational wave memory in an asymptotically flat spacetime; however, the overall magnitude of the memory effect is enhanced by a redshift-dependent factor.  In addition, we find the memory can be affected by lensing.  

\end{abstract}



\maketitle
\allowdisplaybreaks[4]

\section{Introduction}
Gravitational wave memory, a permanent displacement of the gravitational wave detector after the wave has passed, has been known since the work of Zel'dovich and Polnarev~\cite{zeldovich}, extended to the full nonlinear theory of general relativity by Christodoulou\cite{christodoulou} and treated by several authors~\cite{braginsky,damourbl,will,lydia1,lydia2,nullfluid,flatmemory,tolwal1,winicour,strominger,flanagan,favata,cosmo1,twcosmo}. 
See also heuristic ideas in a weak field but no mentioning of memory \cite{turner,epstein}. 
It has been shown \cite{flatmemory} that the memory found by Zel'dovich and Polnarev in a linearized situation and the one by Christodoulou in the nonlinear theory are two different effects, the former (i.e. linear) called ordinary and the latter (i.e. nonlinear) called null memory. The ordinary memory is very small, whereas the null memory is large enough to be detected by Advanced LIGO and other experiments. 
Most of these works treat memory in an asymptotically flat spacetime.  However, we live in an expanding universe, not an asymptotically flat spacetime. Furthermore, the sources of gravitational waves are so rare that the ones that have been detected so far~\cite{ligodetect1,ligodetect2,ligodetect3} have been at distances at which the  expansion of the universe cannot be neglected.  As the detectors become ever more sensitive, one can expect detections due to sources at even greater distances where the expansion of the universe will be even more important.  

A proper treatment of memory in an expanding universe is thus crucial.  This has been done in~\cite{cosmo1} for deSitter spacetime, and in~\cite{twcosmo} for a more general cosmology, but with a particular idealized source.  Our universe, however, has both ordinary and dark matter (which for cosmological purposes can be treated as pressureless dust), as well as dark energy (which can be treated as a cosmological constant).  Although the universe started as a small perturbation of a Friedman-Lema\^itre-Robertson-Walker (FLRW) spacetime, by the present time these perturbations have grown to the point that the local density is very far from that of the unperturbed cosmology.  Thus a realistic treatment of memory in our universe must treat the propagation of gravitational waves through this highly inhomogeneous medium.  Furthermore, unlike the case of deSitter spacetime, the dust equations of motion are coupled to those of the gravitational waves, so a consistent treatment of the gravitational waves should take this coupling into account.  

To treat these complications we take the following approach. We begin by dividing the region far from the source into a ``wave zone'' and a ``cosmological zone.''  Here the wave zone is taken to be the region where the distance from the source is large compared to the wavelength of the waves, but small compared to the Hubble distance.  In the wave zone, the fields that give rise to memory will behave to an excellent approximation just as they do in Minkowski spacetime.  Thus the behavior of the fields is essentially as given in~\cite{flatmemory}. The cosmological zone consists of regions where the distance is not small compared to the Hubble distance.  

If we can determine how the waves change as they propagate from the wave zone to the cosmological zone, then this change, along with the results of~\cite{flatmemory}, will allow us to read off the behavior of memory in the expanding universe.  To treat this propagation in the cosmological zone, we will use the fact that the wavelength of the gravitational waves is short compared to all other length scales in the problem.  We will therefore use the short wavelength approximation of~\cite{yvonne}, though generalized from the vacuum case to the case with dust and a cosmological constant.  This approach is similar to that of~\cite{thorne,isaacson1}, though explicitly taking into account the matter equations of motion and their coupling to gravity.  

In this article, we derive the gravitational wave memory in both the wave zone and the cosmological zone. 
We show that in the wave zone the memory is given via an expression involving the radiated energy per unit solid angle. As the background in the wave zone is approximated well by the Minkowski metric, this memory is computed as in \cite{flatmemory}. We also show that in the cosmological zone the memory is given by the memory computed for the wave zone multiplied by $(1 + z)M$, where $z$ denotes the redshift and $M$ is a magnification factor due to lensing and the Sachs-Wolfe effect. The computations are done with respect to the luminosity distance, which in FLRW spacetimes is the natural replacement for the $r$ coordinate, recalling that in Minkowski spacetime the luminosity distance is equal to $r$. 

Our derivation of the $\Lambda$CDM memory makes use of the results for asymptotically flat spacetimes \cite{flatmemory} by the first and second of the present authors and the short wavelength approximation \cite{yvonne} by Choquet-Bruhat. 
We recall that in \cite{flatmemory} by a gauge-invariant method of perturbing the Weyl tensor away from a Minkwoski background two types of memories are computed, namely the null memory or Christodoulou memory and the ordinary memory that dates back to Zel'dovich and Polnarev. A memory tensor is derived that consists of exactly these two parts. The null memory, which is much larger than the (tiny) ordinary one, is due to energy radiated away to infinity per unit solid angle whereas the ordinary memory is due to changes in the $(r,r)$ component of the electric part of the Weyl tensor. Decomposing the memory into spherical harmonics it is shown that the major part of the null memory is due to energy radiated in the $l=2$ modes. This paper restricts attention to the null memory, i.e. Christodoulou memory. However, in cosmological spacetimes there is no ``null infinity'', which plays a crucial role in analyzing radiation and memory in asymptotically Minkowskian spacetimes. The solution to this problem is provided first by our separate treatment of memory in the wave and cosmological zones and by using the short wavelength approximation. The latter \cite{yvonne} makes use of the fact that the wavelength of the gravitational waves is short compared to all other scales in the problem. In particular, we consider a one-parameter family of spacetime metrics consisting of a background metric plus $\omega^{-2}$ times a radiative metric of frequency $\omega$. The short wavelength limit is then the limit for large $\omega$. We start with the background to be FLRW on large scales and add a perturbation that takes into account the local inhomogeneities in our universe. 
We also express each component of the stress-energy tensor in a similar way. 
Thus, the spacetime metric and matter content are provided by the tensorfields given in (\ref{gwave})-(\ref{uwave}). 
In situations as studied here, our spacetime metric solves the Einstein equations asymptotically in the high frequency limit, that is to a given order in $1/{\omega}$. 
An interesting feature analyzed is that these perturbations can only be purely gravitational or purely fluid. Therefore, doing a gravitational perturbation, the fluid part vanishes at lowest order. It then follows that the decay behavior of the gravitational wave amplitude is given by a simple argument. In particular, apart from gauge terms, it is computed using the divergence of the null geodesic vector field introduced in the next section. 
The inhomogeneities in our universe generate curvature that can interfere with the waves. As these are traveling on 
null geodesics, we investigate the latter and seek to understand how the Weyl curvature changes. As light rays follow null geodesics, we explore what is known about gravitational lensing in the corresponding situation to show that the Weyl tensor is multiplied by a magnification factor due to gravitational lensing. Hereby, we use results obtained in \cite{Laguna:2009re}. 
Finally, knowing how the Weyl curvature behaves, we use this in the geodesic deviation equation to compute the memory in the two different zones, and thereby derive the results mentioned above.

The remainder of this paper is organized as follows. Section~\ref{sec:short-waves} will treat the short wavelength cosmological gravitational waves and will obtain a result for the behavior of the Weyl tensor as the waves propagate from the wave zone to the cosmological zone.  Section~\ref{sec:cosmological-memory} will use the results of Sec.~\ref{sec:short-waves} to obtain the cosmological memory.  Section~\ref{sec:astro-imply} contains our conclusions and the observational implication of our work.    

\section{Gravitational Waves in the Short Wavelength Approximation}
\label{sec:short-waves}

\subsection{Field Equations}

We want to consider waves in a cosmology that consists of dust and a cosmological constant.  However, we do not want to assume that the spacetime is nearly FLRW, since at the present time initially small density perturbations have become large.  Instead, we will use the approximation that the wavelength of the waves is short compared to all other scales in the problem, and we will use the weak progressive wave method of~\cite{yvonne}.

We begin with a background solution of the Einstein field equations with dust and a cosmological constant.  This background solution consists 
of a metric ${{\bar g}_{ab}}({x^\mu})$, a dust density ${\bar \rho}({x^\mu})$ and a four-velocity ${{\bar u}_a}({x^\mu})$ that satisfy 
\be
{{\bar R}_{ab}} - {\textstyle {1 \over 2}} {\bar R} {{\bar g}_{ab}} + \Lambda {{\bar g}_{ab}} - 8 \pi \bar\rho \; {{\bar u}_a}{{\bar u}_b} = 0 \; \; \; .
\label{efe0}
\ee
Here ${\bar R}_{ab}$ is the Ricci tensor of ${\bar g}_{ab}$, $\bar R$ is the scalar curvature and $\Lambda$ is the cosmological constant. This background represents the cosmology of our evolving universe, which we take to be FLRW on large scales, though with (possibly large) density contrasts on small scales. 

This background, however, does not describe gravitational waves or their sources.  With this in mind, we introduce another one-parameter family of tensor fields ${{\hat g}_{ab}} ({x^\mu},\xi), \, {\hat \rho} ({x^\mu},\xi), \, {{\hat u}_a} ({x^\mu},\xi)$, and a scalar field $\phi ({x^\mu})$. These perturbations represent high-frequency deformations of the background that are uniformly bounded in $\xi$, with the only restriction that the length scale of the inhomogeneities is large compared to the much smaller wavelength of the gravitational waves. 

The full spacetime and matter content of the universe is then given by the one-parameter family of tensor fields $({g_{ab}}, \rho, {u_a})$, which we write as 
\bea
{g_{ab}} &=& {{\bar g}_{ab}} ({x^\mu}) + {\omega ^{-2}}{{\hat g}_{ab}}({x^\mu},\omega \phi ({x^\mu})) \,.
\label{gwave}
\\
\rho &=& {\bar \rho}({x^\mu}) + {\omega ^{-1}} {\hat \rho} ({x^\mu},\omega \phi({x^\mu})) \,,
\label{rhowave}
\\
{u_a} &=& {{\bar u}_a}({x^\mu}) + {\omega ^{-1}}  {{\hat u}_a} ({x^\mu},\omega \phi({x^\mu})) \,,
\label{uwave}
\eea
where $\omega$ is the frequency of the perturbations. The fields $({g_{ab}}, \rho, {u_a})$ represent our universe in the sense that they satisfy the Einstein-fluid equations \emph{to the appropriate order}:
\be
{R_{ab}} - {\textstyle {1 \over 2}} R {g_{ab}} + \Lambda {g_{ab}} - 8 \pi \rho \; {u_a}{u_b} =  O (
{\omega ^{-2}}) \; \; \;,
\label{efe}
\ee
where ${R}_{ab}$ is the Ricci tensor of ${g}_{ab}$, and $R$ is the scalar curvature. Note that here the parameter $\omega$ plays a dual role, both as the frequency of the perturbation and as an inverse amplitude.  The  surfaces $\phi= {\rm const}$ are wavefronts, since in the large $\omega$ limit the waves vary rapidly in the direction perpendicular to them. Note also that Eq.~(\ref{efe}) describes the waves only in the region away from their sources.

This approach is somewhat different from the usual perturbative approach in general relativity.  In the usual perturbative approach, we assume that there is a one-parameter family of metric tensor fields, where each member of the family is an exact solution of the field equations, but we only calculate that family to first order in the parameter.  In contrast, in the weak progressive wave approach, the one-parameter family of metric tensor fields is not expanded only to first order in the parameter, but rather the field equations themselves, (Eq.~(\ref{efe})) are only satisfied to a given order in $\omega ^{-1}$. The approach is similar to that of Isaacson in~\cite{isaacson1}, but differs from that of Isaacson in \cite{isaacson2} where the waves can be strong enough that their effective stress-energy has a strong effect on the background geometry.  The factor of $\omega^{-2}$ in Eq.~(\ref{gwave}) and conditions entailed by Eq.~(\ref{efe}) ensure that no such strong effect is present.  Our approach will also differ from the usual perturbative approach in that we never use the diffeomorphism invariance of the theory to choose a particular gauge.  Instead, in keeping with the methods of \cite{flatmemory,cosmo1} all our calculations and results are stated in a way that is manifestly gauge invariant. 

We now compute the covariant derivative operator and Riemann tensor of $g_{ab}$.  For any one-form $A_a$, we have exactly that \cite{wald} 
\be
{\nabla _a} {A_b} = {{\bar \nabla}_a}{A_b} - {C^c _{ab}}{A_c} \,,
\ee
where ${\nabla}_a$ and ${\bar \nabla}_a$ are the covariant derivative operators of $g_{ab}$ and ${\bar g}_{ab}$ respectively, and where the difference tensor $C^c _{ab}$ is given by\footnote{The formula for the difference tensor $C^c _{ab}$ is similar to that of the Christoffel symbol $\Gamma ^c _{ab}$ and indeed the Christoffel symbol of a metric $g_{ab}$ is the difference tensor between the covariant derivative $\nabla _a$ and the coordinate derivative $\partial _a$.  See \cite{wald} for details.}
\be
{C^c _{ab}} = {\textstyle {1 \over 2}} {g^{cd}} \left ( {{\bar \nabla}_a}{g_{bd}} + {{\bar \nabla}_b}{g_{ad}} -
{{\bar \nabla}_d}{g_{ab}} \right ) \,.
\label{Cdef}
\ee
Using Eq.~(\ref{gwave}) in Eq.~(\ref{Cdef}) and expanding in $1/\omega$, we then obtain 
\begin{align}
{C^c _{ab}} &= {\textstyle {1 \over 2}} {{\bar g}^{cd}} \left [ {\omega ^{-1}} ( {k_a} {{{\hat g}'}_{bd}} + {k_b} {{{\hat g}'}_{ad}} - {k_d} {{{\hat g}'}_{ab}} ) \right.
\nonumber \\
&\left. + {\omega ^{-2}} ( {{\bar \nabla}_a} {{\hat g}_{bd}} + {{\bar \nabla}_b} {{\hat g}_{ad}} - {{\bar \nabla}_d} {{\hat g}_{ab}} ) \right ] + O ({\omega ^{-3}}) \,,
\label{Cwave}
\end{align}
where $k_a = {\nabla _a}\phi$ and ${\hat g}_{ab}$ is considered as a function of $x^\mu$ and $\xi$.  A prime denotes derivative with respect to $\xi$ and ${\bar \nabla}_a$ takes derivatives only with respect to $x^\mu$.  It is only once all these operations are performed that we evaluate all quantities at $\xi = \omega \phi({x^\mu})$.  

Let us now  compute the Riemann tensor to $O(\omega ^{-1})$.  A standard result of general relativity is that the Riemann tensor of a 
metric $g_{ab}$ can be written exactly as
\be
{{R_{abc}}^d} = {\bar{R}_{abc}{}^d} + {{\bar \nabla }_b} {C^d _{ac}} - {{\bar \nabla }_a} {C^d _{bc}} 
+ {C^e _{ac}}{C^d _{be}} - {C^e _{bc}} {C^d _{ae}} \,,
\label{Rdef}
\ee
where $\bar{R}_{abc}{}^d$ is the Riemann tensor of the background.  From Eq.~(\ref{Cwave}), it follows that terms quadratic in $C^c _{ab}$ are $O(\omega^{-2})$.  Using Eq.~(\ref{Cwave}) in Eq.~(\ref{Rdef}), we obtain 
\be
{{R_{abc}}^d} =  {\bar{R}_{abc}{}^d} + {R^{(0)}_{abc}{}^d} + {\omega ^{-1}} {R^{(1)}_{abc}{}^d} + O({\omega ^{-2}})
\ee
where the tensors $R^{(0)}_{abc}{}^d$ and $R^{(1)}_{abc}{}^d$ are given by
\bea
{R^{(0)}_{abc}{}^d} &=&  {{\bar g}^{de}} \left [ {k_b}{k_{[c}}{{{\hat g}''}_{e]a}} - {k_a}{k_{[c}}{{{\hat g}''}_{e]b}}
 \right ]
\\
{R^{(1)}_{abc}{}^d} &=&   {{\bar g}^{de}} \bigl [ {{{\hat g}'}_{e[a}}{{\bar \nabla}_{b]}}{k_c} -  {{{\hat g}'}_{c[a}}{{\bar \nabla}_{b]}}{k_e} 
\nonumber
\\
&+& {k_{[b}}{{\bar \nabla}_{a]}}{{{\hat g}'}_{ce}} + {k_b}{{\bar \nabla}_{[c}}{{{\hat g}'}_{e]a}}
- {k_a}{{\bar \nabla}_{[c}}{{{\hat g}'}_{e]b}}
\bigr ]\,.
\label{Rwave}
\eea
Contracting on indices $b$ and $d$ we find that the Ricci tensor of $g_{ab}$ takes the form
\be
{R_{ac}} = {{\bar R}_{ac}} + {R^{(0)} _{ac}} + {\omega ^{-1}} {R^{(1)} _{ac}} + O({\omega ^{-2}}) \,.
\ee
Here ${\bar R}_{ac}$ is the Ricci tensor of the background, and the tensors ${R^{(0)} _{ac}}$ and ${R^{(1)} _{ac}}$ are given by
\begin{align}
{R^{(0)} _{ac}} &= - {\textstyle {\frac 1 2}} {k^b}{k_b} {{{\hat g}''}_{ac}} + {k_{(a}}{{P''}_{c)}} \,,
\label{Ric0}
\\
{R^{(1)} _{ac}} &= - {k^b}{{\bar \nabla }_b}{{{\hat g}'}_{ac}} - {\textstyle {\frac 1 2}} ( {{\bar \nabla}_b}{k^b}) 
{{{\hat g}'}_{ac}} + {{\bar \nabla }_{(a}}{{P'}_{c)}} + {k_{(a}}{L_{c)}} \,,
\end{align}
where we raise and lower indices with the background metric, and the one-forms $P_a$ and $L_a$ are given by
\begin{align}
{P_a} &= {{\hat g}_{ab}}{k^b} - {\textstyle {\frac 1 2}} {k_a}{\hat g}\,,
\\
{L_a} &= {{\bar \nabla}_b}{{{\hat g}'}_a {} ^b} - {\textstyle {\frac 1 2}} {{\bar \nabla}_a}{\hat g}\,,
\end{align}
with ${\hat g} = {{{\hat g}_a}^a}$. The quantities $P_a$ and $L_a$ can be changed by a gauge transformation, and indeed in the usual perturbative approach the radiation gauge is used to set analogous quantities to zero. As stated before, we take a gauge-agnostic view here, and thus, we will not impose such gauge conditions. 

\subsection{Solutions to $O(\omega^{0})$}

We now consider the consequences of the field equations, Eq.~(\ref{efe}). Equations~(\ref{rhowave}) and (\ref{uwave}) imply the matter terms give no corrections to the field equations at ${\cal{O}}(\omega^0)$. Therefore, Eq.~(\ref{efe}) implies that ${R^{(0)} _{ab}} = 0$.  Thus at zeroth order our results are the same as those of~\cite{yvonne} for vacuum progressive waves.  These results are comprised of two cases: 

\begin{itemize}
\item {\bf{Case (i): ${k^a}{k_a} =0$}}.
\vspace{0.2cm}
\\
It then follows from Eq.~(\ref{Ric0}) that ${{P''}_a}=0$, but then we must have ${{P'}_a} = 0$, since otherwise ${P}_a$ would grow linearly with $\xi$, and this would violate our assumption that ${\hat g}_{ab}$ is uniformly bounded. Since ${k_a} = {{\bar \nabla }_a} \phi $, it follows from ${k^a}{k_a}=0$ that ${k^a}{{\bar \nabla}_a}{k^b}=0$, i.e.~the waves propagate along null geodesics. 
\item {\bf{Case (ii): ${k^a}{k_a} \ne 0$}}.
\vspace{0.2cm}
\\ 
It then follows from Eq.~(\ref{Ric0}) that ${\hat g}_{ab}$ takes the form
${{\hat g}_{ab}} = {k_{(a}}{s_{b)}}$ for some $s_a$, but this is a pure gauge mode. Consider a vector field ${\eta _a}= {\omega ^{-3}} {q_a}({x^\mu},\omega \phi)$ and act on the background metric ${\bar g}_{ab}$ with a diffeomorphism along $\eta ^a$. This produces a physically identical metric that, to ${\cal{O}}(\omega^{-2})$, takes the form
\be
{{\bar g}_{ab}} + {{\cal L}_\eta}{{\bar g}_{ab}} = {{\bar g}_{ab}} + {\omega ^{-2}} {k_{(a}}{{q'}_{b)}} \,.
\ee
This is a ${\hat g}_{ab}$ metric of the form ${{\hat g}_{ab}}={k_{(a}}{{s}_{b)}}$, which is thus a pure gauge mode.  
\vspace{0.2cm}
\\
For our purposes, a different way of seeing that a perturbation of the form ${{\hat g}_{ab}} = {k_{(a}}{s_{b)}}$ is pure gauge is to note that from Eq.~(\ref{Rwave}) and the vanishing of ${R^{(0)} _{ab}}$ it follows that to $O(\omega^{0})$ the Weyl tensor is 
${{\bar C}_{abcd}} + {C^{(0)}_{abcd}}$ where ${\bar C}_{abcd}$ is the Weyl tensor of the background
and ${C^{(0)}_{abcd}}$ is the zeroth order Weyl tensor of the wave, which is given by
\be
{C^{(0)}_{abcd}} =  \left [ {k_c}{k_{[b}}{{{\hat g}''}_{a]d}} - {k_d}{k_{[b}}{{{\hat g}''}_{a]c}}
 \right ] \,.
\label{Weyl0}
\ee
Note then that a ${\hat g}_{ab}$ of the form 
${{\hat g}_{ab}} = {k_{(a}}{s_{b)}}$ leads to a zero Weyl tensor of the wave.
\end{itemize}

\subsection{Solutions to $O(\omega^{-1})$}

Let us begin by considering the equation of motion for the matter fields.  From the Bianchi identities and Eq.~(\ref{efe}) we obtain
\be
{\nabla ^a}(\rho \; {u_a}{u_b}) = O ({\omega ^{-1}}) \,,
\ee
from which it follows that
\bea
{u^a}{\nabla _a}{u_b} = O ({\omega ^{-1}}) \,,
\label{euler1}
\\
{u^a}{\nabla _a} \rho + \rho {\nabla _a}{u^a} = O ({\omega ^{-1}})\,.
\label{euler2}
\eea
Now using Eqs.~(\ref{rhowave}) and (\ref{uwave}) in Eqs.~(\ref{euler1}) and (\ref{euler2}), we find that to $O(\omega^{0})$
\begin{align}
{k^a}{{\bar u}_a} {{\hat u}_b} = 0 \,,
\label{fluid1}
\\
{k^a}{{\bar u}_a}{\hat \rho} + {\bar \rho}{k^a} {{\hat u}_a}=0 \,.
\label{fluid2}
\end{align}
From Eq.~(\ref{fluid1}) it follows that either ${k^a}{{\bar u}_a}=0$ or $ {{\hat u}_a} = 0$.  However, ${k^a}{{\bar u}_a}=0$ is not compatible with ${k^a}{k_a}=0$ (if $k^a$ is orthogonal to ${\bar u}^a$, then $k^a$ is spacelike and therefore cannot be null).  Thus, if we have a (not pure gauge) gravitational wave, then we must have ${{\hat u}_a} = 0$.  It then follows from (\ref{fluid2}) that ${\hat \rho } = 0$.  In other words, the fluid perturbation vanishes.  In physical terms, what all this means is that gravitational perturbations (which travel at the speed of light) and fluid perturbations (which travel at the speed of sound, in this case zero because the fluid is dust) cannot have the same wavevector.  Thus a perturbation with a single wavevector must be pure gravity or pure fluid.  

Let us now consider a non-trivial gravitational perturbation. Since the fluid perturbation vanishes at lowest order, it follows that ${R^{(1)} _{ab}}=0$. That is, even to $O(\omega^{-1})$ the field equations reduce to that of vacuum.  From ${R^{(1)} _{ab}}=0$ and ${{P'}_a}=0$ we obtain
\be
- {k^b}{{\bar \nabla }_b}{{{\hat g}'}_{ac}} - {\textstyle {\frac 1 2}} ( {{\bar \nabla}_b}{k^b}) 
{{{\hat g}'}_{ac}}  + {k_{(a}}{L_{c)}} = 0 \,.
\label{kgradv}
\ee  
That is, up to terms that are pure gauge, the fall-off of the gravitational wave amplitude is determined by the properties of the divergence of the null geodesic vector field $k^{a}$. This result can be stated in a manifestly gauge invariant way as follows. Taking the derivative with respect to $\xi$ of Eq.~(\ref{kgradv}) and using the result in Eq.~(\ref{Weyl0}) we obtain  
\be
{k^e}{{\bar \nabla_e}}{C^{(0)}_{abcd}} = - {\textstyle {\frac 1 2}} ( {{\bar \nabla}_e}{k^e}) {C^{(0)}_{abcd}}
\,.
\label{kgradWeyl}
\ee

\subsection{Implications in Homogeneous Background Spacetimes}

We now consider the implications of Eq.~(\ref{kgradWeyl}) in Minkowski spacetime and in an FLRW spacetime.  Recall that the line element of Minkowski spacetime can be written in spherical polar coordinates as 
\be
d {s^2} = - d {t^2} + d {r^2} + {r^2} ( d {\theta ^2} + {\sin ^2} \theta d {\varphi ^2} ) \,. 
\ee
It then follows that $k_a$ (defined to be the affinely parameterized radial outgoing null geodesic) is given by 
${k_a}={\nabla _a}(r-t)$ and therefore that ${\nabla _a}{k^a} = 2/r$.  It then follows from Eq.~(\ref{kgradWeyl}) that 
\be
{k^e}{\nabla _e} \left ( r \; {C^{(0)}_{abcd}}\right ) = 0 \,.
\label{flatpeel}
\ee 
Similarly, in FLRW spacetime the line element is 
\be
d {s^2} = - d {t^2} + {a^2}(t) \left [ d {r^2} + {r^2} ( d {\theta ^2} + {\sin ^2} \theta d {\varphi ^2} ) \right ]
\,,
\label{gFLRW}
\ee  
and it then follows that 
\be
{k_a}={\nabla _a}r - {a^{-1}}{\nabla _a}t
\label{kaFLRW}
\ee
and therefore, that 
\be
{\nabla _a}{k^a} = {\frac 2 {a^2}} \left ( {\dot a}  + {\frac 1 r} \right ) \,.
\label{divkFLRW1}
\ee
However, it follows from Eqs.~(\ref{gFLRW}) and (\ref{kaFLRW}) that ${k^a}{\nabla _a}r = {a^{-2}}$ and that ${k^a}{\nabla _a}a= {\dot a}/a$, and thus from Eq.~(\ref{divkFLRW1}) we know that
\be
{\nabla _a}{k^a} = {\frac 2 {a r}} {k^a}{\nabla _a} (a r) \,.
\label{divkFLRW2}
\ee
From Eq.~(\ref{kgradWeyl}) it then follows that 
\be
{k^e}{\nabla _e} \left ( a \; r \; {{C^{(0)}}_{abcd}}\right ) = 0 \,.
\label{FLRWpeel}
\ee

\subsection{Implications in an Inhomogeneous Background Spacetimes}
\label{sec:inhomo}

The background spacetimes we consider are FLRW on large scales, but on small scales the null geodesics can encounter curvature that can lead to modifications in wave propagation.  How, then are we to take into account this additional effect? Since light rays are described by null geodesics, the effect of lensing on the brightness of a light wave is given by an equation of the same form as Eq.~(\ref{kgradWeyl}).  We therefore expect that the additional effect of inhomogeneities is precisely to multiply the Weyl tensor by a \emph{magnification factor} due to gravitational lensing. In this section, we will re-derive this result, using results from~\cite{Laguna:2009re}. 

Consider a scalar ${\cal{A}}$ that satisfies the equation
\be
\label{A-eq}
k^{e} \bar\nabla_{e} {\cal{A}} = - \frac{1}{2} {\cal{A}} \bar\nabla_{e} k^{e}\,.
\ee
We can then use this equation to rewrite Eq.~\eqref{kgradWeyl} as
\be
{k^e}{{\bar \nabla_e}}{C^{(0)}_{abcd}} - {\cal{A}}^{-1} C^{(0)}_{abcd} k^{e} \bar\nabla_{e} {\cal{A}} = 0\,.
\ee
We recognize this expression as the product-rule distributed version of 
\be
k^{e} \bar\nabla_{e} \left( {\cal{A}}^{-1} C^{(0)}_{abcd}\right) = 0\,.
\ee 

The expression above can be used to calculate the memory, if we can find an expression for ${\cal{A}}^{-1}$. 
Let us thus specialize to an inhomogeneous FLRW spacetime background, described by the line element 
\begin{align}
d {s^2} &= - \left(1 + 2 \Phi\right) a^{2}(\tau) d \tau^2
+ a^{2}(\tau) \left(1 - 2 \Psi\right) \delta_{ij} dx^{i} dx^{j}\,, 
\label{gFLRW-perturbed}
\end{align}
where $\delta_{ij}$ is the Kronecker delta and $(\Phi,\Psi)$ are matter inhomogeneities that in principle depend on conformal time $\tau$
(related to the time coordinate $t$ via $dt = a(\tau) d\tau$) and the Cartesian coordinates $x^{i}$. 
Such a perturbed FLRW spacetime suggests similar perturbative decompositions of other quantities, 
such as the scalar function ${\cal{A}} = {\cal{A}}_{0} \left(1 + \zeta\right)$, where ${\cal{A}}_{0}$ and $\zeta$
are independent and linearly-dependent on the matter inhomogeneities respectively. By comparison with
Eq.~\eqref{FLRWpeel}, we immediately see that ${\cal{A}}_{0} = 1/(a r)$. 

The part of ${\cal{A}}$ that is linearly proportional to the matter inhomogeneities can be obtained by solving 
Eq.~\eqref{A-eq} linearized in $(\Phi,\Psi)$. This equation, in turn, depends on the solution to the null-geodesic equation 
in the perturbed spacetime of Eq.~\eqref{gFLRW-perturbed}. The calculation, using the methods of \cite{Laguna:2009re}, are given in appendix B.  Here, we just present the final result:
\bea
\zeta &=& \Psi - {\Psi _e} 
\nonumber
\\
&+& {\frac 1 2} {\int _0 ^\lambda } \; {\frac {d {\lambda '}} {{({\lambda '})}^2}} \; 
{\int _0 ^{\lambda '} } \;  d {\lambda ''} \; {D_A}{D^A}(\Phi + \Psi)  \; \; \; .
\label{xirecap}
\eea
where $\lambda$ is the affine parameter of null geodesics in the non-expanding but inhomogeneous spacetime (Eq.~\eqref{gFLRW-perturbed}
with $a(\tau)$ set to unity), $\Psi_{e}$ is the value of this $\Psi$ at emission, and ${D_A}{D^A}$ is the laplacian on the unit two sphere. The first term in the above equation corresponds to the standard Sachs-Wolfe effect~\cite{Sachs:1967er}, while the second
is a magnification due to lensing~\cite{Laguna:2009re}.  The result in eqn. (\ref{xirecap}) is similar to the corresponding equation in  
\cite{Laguna:2009re}; however, we correct an overall minus sign in that reference, and we improve the accuracy of the terms involving angular derivatives.  

With $\cal A$ calculated, we then find that Eq.~\eqref{kgradWeyl} in the spacetime of Eq.~\eqref{gFLRW-perturbed} simplifies to
\be
k^{f} \bar{\nabla}_{f} \left[C^{(0)}_{bcde} a r \left(1 - \zeta \right)\right] = 0\,,
\label{FLRWpeel-perturbed}
\ee
where once more we have expanded in small matter inhomogeneities. As we will see in the next section, the $\zeta$ term magnifies the signal,
thus magnifying the memory effect.  

\section{Cosmological Memory}
\label{sec:cosmological-memory}

We now apply the results of the previous section, and in particular of Eq.~(\ref{FLRWpeel}), to gravitational wave
memory.  Let us begin by introducing two 4-dimensional spacetime regions: the \emph{wave zone} and the \emph{cosmological zone}. The wave zone is defined through the asymptotic relation $H_{0}^{-1} \gg r \gg \lambda$, while the cosmological zone is defined through $r \gtrsim H_{0}^{-1}$, where $r$ is the distance from the gravitational wave emitting source to a field point, $\lambda$ is the gravitational wave wavelength and $H_{0}$ is the Hubble parameter today. In the wave zone, the FLRW background spacetime can be well-approximated by the Minkowski metric, while in the cosmological zone one must use the full FLRW metric. Let us then imagine that a gravitational wave is emitted at $r_{0}=0$, detected first at $r_{1}$ in the wave zone and then detected again at $r_{2}$ in the cosmological zone\footnote{Note that these conventions are opposite to those used in~\cite{Laguna:2009re}, where the observer is at the \emph{end} of the gravitational wave worldline. This, for example, affects the sign of the $n^{i}$ vector, which should point in the direction of arrival of the gravitational wave. However, since only even powers of $n^{i}$ enter our equations, the signals cancel and the results are the same.}. The goal of this section is to compare a memory measurement in the wave zone to another measurement in the cosmological zone. We begin by considering the cosmological memory in a homogenous FLRW background, and then conclude this section with a discussion of the effect of inhomogeneities. 

Let us first recall some of the basic properties of the gravitational wave memory~\cite{flatmemory}. For two nearby geodesics with four-velocity $u^a$ and separation $s^a$ acted on by a gravitational wave with Weyl tensor $C_{abcd}$, the geodesic deviation equation requires that 
\be
{{\ddot s}^a} = - {{C^a} _{bcd}}{u^b}{s^c}{u^d}
\label{geodev}
\ee 
where an overdot denotes derivative with respect to the proper time of the geodesics. For simplicity we assume an initial displacement orthogonal to the direction of propagation of the wave, and we consider only the memory due to energy radiated to infinity~\cite{christodoulou}; we use capital letters to denote indices in this two-sphere of orthogonal directions. 

Measurements in the wave zone can be related to measurements in the cosmological zone through the definition of the luminosity distance. In Minkowski spacetime, the luminosity distance is the same as the usual $r$ coordinate, but in an FLRW spacetime this is not the case, since the FLRW $r$ coordinate does not have a physical meaning by itself. The luminosity distance is defined as $d_{L} = [P/(4 \pi F)]^{1/2}$, where $P$ is the power of a light source and $F$ is the flux through a sphere of radius equal to the luminosity distance. Because the time of flight and the frequency of photons and gravitons redshifts as they propagate in an expanding universe, we then have that $d_{L} = r a (1 + z)$, where $z$ is the redshift and $a$ is the scale factor at the location of the measurement. For the general treatment we present in this paper, we will find it convenient to express all of our results in terms of $d_{L}$.   

Equation~(\ref{geodev}) implies that after the wave has passed there will be a residual change in the separation $\Delta {s^a}$.  Let the original separation $s$ be in the $B$ direction.  Then the change in separation $\Delta s$ in the $A$ direction is given by
\be
\Delta s = - {\frac s {d_L}} {{m^A}_B}
\ee    
where the memory tensor ${{m^A}_B}$ is given by
\be
{{m^A}_B} = {\int _{-\infty} ^\infty} d {\tilde t} {\int _{-\infty} ^{\tilde t}} d t \; ({d_L} {C_{abcd}}{x^a}{u^b}{y^c}{u^d}) \; \; \; .
\label{memtensor}
\ee
Here $x^a$ and $y^a$ are respectively unit vectors in the $A$ and $B$ directions.  For simplicity, we treat the case where $x^a$ and $y^a$ are orthogonal to the direction of wave propagation. One can write a similar expression for the differential change in arm length given a gravitational wave at a generic sky location through the inclusion of an antenna pattern tensor, as we will show in appendix A. 

In a spacetime with a Minkowski background, there is a relation between the memory tensor $m_{AB}$ and the energy radiated to null infinity.  Specifically, let $F(\theta,\varphi)$ be the energy per unit
solid angle radiated to infinity in the direction given by the two-sphere coordinates $(\theta,\varphi)$ and let $D_A$ be the derivative operator on the unit two-sphere.  Then $m_{AB}$ is the unique traceless tensor satisfying
\be
{D^A}{m_{AB}} = {D_B}\Phi
\label{divm}
\ee
where $\Phi$ is the solution of
\be
{D^A}{D_A} \Phi = - 8 \pi (F-{F_{[1]}})
\label{LaplacePhi}
\ee
and $F_{[1]}$ is the sum of the $\ell =0$ and $\ell =1$ pieces of $F$. 

Equations~(\ref{geodev}-\ref{memtensor}) are general and thus apply both at $r_1$ with a memory tensor $m^{(1)}_{AB}$ and at $r_2$ with a memory tensor $m^{(2)}_{AB}$. Equations~(\ref{divm}-\ref{LaplacePhi}), on the other hand, are specific to spacetimes with a Minkowski background and thus apply only at $r_1$.  Thus, our strategy for calculating cosmological memory is as follows: $m^{(1)}_{AB}$ is determined by the local (i.e.~at $r_1$) $F$ using the usual Minkowski spacetime methods.  Then, using Eqs.~(\ref{FLRWpeel}) and (\ref{memtensor}) we will determine a relation between $m^{(2)}_{AB}$
and $m^{(1)}_{AB}$, and thus, allow the determination of $m^{(2)}_{AB}$ from the local $F$.  

Let ${w^a},\, {z^a}$ and $q^a$ be the vectors that start as ${u^a},\, {x^a}$ and $y^a$ respectively at $r_1$ and are parallel propagated along $k^a$ toward $r_2$. 
With these definitions in hand, it then follows from Eq.~(\ref{FLRWpeel}) that the quantity 
\be
a \; r  \; {C^{(0)}_{abcd}}{z^a}{w^b}{q^c}{w^d} = {\rm constant}
\label{Eab1}
\ee
along the null geodesic to which $k^a$ is tangent, and therefore this quantity is the same at $r_2$ as it is at $r_1$.  However, it follows from the
properties of FLRW spacetimes that ${z^a}={x^a}$ and ${q^a}={y^a}$ and 
\be
{w^a} = {\frac 1 {2 {a_1}}} \left [ 2 a {u^a} + ({a_1 ^2} - {a^2}) {k^a} \right ]
\label{wa}
\ee 
where $a_1 = a(t_1)$ and $t_1$ is the time at which the gravitational waves cross $r_1$, which for our purposes is essentially the time the waves are emitted.  However, it follows from Eq.~(\ref{Weyl0}) that ${k^a}{{C^{(0)}}_{abcd}}=0$, and then from Eqs.~(\ref{Eab1}) and (\ref{wa}) that
\be
{a^3} \; r \;  {C^{(0)}_{abcd}} \; {x^a}{u^b}{y^c}{u^d} = {\rm constant}
\label{Eab2}
\ee
along the null geodesic to which $k^a$ is tangent, and therefore, this quantity is the same 
at $r_2$ as it is at $r_1$: 
\be
\left({a_{1}^3} \; r_{1} \right) \left.{C^{(0)}_{abcd}} \; {x^a}{u^b}{y^c}{u^d}\right|_{r=r_{1}} 
\!\!\! =
\left({a_{2}^3} \; r_{2} \right) \left.{C^{(0)}_{abcd}} \; {x^a}{u^b}{y^c}{u^d}\right|_{r=r_{2}}\,,
\label{Eab3}
\ee
where $a_{1,2} = a({t_{1,2}})$ and $t_{1,2}$ is the time when the gravitational wave is detected at $r_{1,2}$. 

Let us use this relation to express the memory in terms of the luminosity distance and the redshift. For the gravitational waves of interest to us, the luminosity distance from the source to the wave zone measurement at $r_{1}$ is simply $d_{L}^{(1)} = r_{1} a_{1} (1 + z_{1}) \sim r_{1} a_{1}$, while the luminosity distance from the source to the cosmological measurement at $r_{2}$ is $d_{L}^{(2)} = d_{L}^{(1)} + r_{2} a_{2} (1 + z_{2}) \approx r_{2} a_{2} (1 + z_{2})$, where $z_{1} \ll 1$ is the redshift between $r_{0}$ and $r_{1}$ and $z_{2} = 1- a_{2}/a_{1}$ is the redshift between $r_{0}$ (or $r_{1}$) and $r_{2}$. Thus, it follows from Eq.~(\ref{Eab3}) that
\be
{d_{L2}} \; \left(1 + z_{2}\right) {{\left .  {C^{(0)}_{abcd}} \; {x^a}{u^b}{y^c}{u^d} \right |}_{r={r_2}}} \!\!\!= 
d_{L1}\; {{\left .  {C^{(0)}_{abcd}} \; {x^a}{u^b}{y^c}{u^d} \right |}_{r={r_1}}}
\ee
We can now use this expression in Eq.~(\ref{memtensor}), together with the fact that $dt$ at $r_2$ is $1+z_{2}$ times $dt$ at $r_1$ to find 
\be
{m^{(2)}_{AB}}=(1+z_{2}) {m^{(1)}_{AB}}\,.
\ee  

This {\emph {would}} be the result if the spacetime were exactly FLRW without matter inhomogeneities. As we discovered in Sec.~\ref{sec:inhomo}, matter inhomogeneities introduce a lensing correction to the amplitude of gravitational waves. Following the same reasoning as above and using Eq.~\eqref{FLRWpeel-perturbed}, we then find
\begin{align}
& {d_{L2}} \; (1+z_{2}) (1 - \zeta_{2}) \; {{\left .  {C^{(0)}_{abcd}} \; {x^a}{u^b}{y^c}{u^d} \right |}_{r={r_2}}} = 
\nonumber \\
&
d_{L1} \;  {{\left .  {C^{(0)}_{abcd}} \; {x^a}{u^b}{y^c}{u^d} \right |}_{r={r_1}}}\,,
\end{align}
where the contribution of $\zeta$ at $r_{1}$ vanishes because it is very close to the emission point $\lambda_{e}$. 
Using again Eq.~(\ref{memtensor}) and expanding about $\zeta_{2} \ll 1$, we then find
\be
{m^{(2)}_{AB}}=(1+z_{2}) \left(1 + \zeta_{2}\right) {m^{(1)}_{AB}}
\ee
where $\zeta_{2}$ induces a magnification or a demagnification (analogous to focusing
and de-focusing) of the signal.  This result is consistent with the analysis of~\cite{Laguna:2009re} and the results of~\cite{cosmo1,twcosmo}.  

\section{Astrophysical Implications}
\label{sec:astro-imply}

We have seen that the gravitational wave memory acquires a redshift enhancement and a lensing correction when the gravitational waves travel large cosmological distances through matter inhomogeneities. Let us now study the degree to which these inhomogeneous and cosmological modifications affect astrophysical observations with gravitational waves. 

Current (second-generation), ground-based detectors are sensitive only to low redshift sources. This is because ground-based instruments operate in the hecto-Hz range, allowing for the detection of black hole mergers with masses not larger than ${\cal{O}}(10^{2} M_{\odot})$, which restricts the detection range to redshifts below ${\cal{O}}(10^{-1})$. This immediately implies that the redshift magnification will not exceed of order $10\%$, while lensing modifications are probably an order of magnitude smaller than that. Such small modifications will not magnify the gravitational wave memory enough to make it detectable with single observations. Recent work has suggested that the stacking of multiple observations may make the memory effect detectable~\cite{Lasky:2016knh}, and here including the redshift magnification will probably be important, although lensing is unlikely to matter. Fortunately, the waveform models that the LIGO collaboration uses already include the redshift magnification, and thus, no modifications to the analysis are needed. 

Once the next generation (third-generation) gravitational wave detectors come online, the redshift enhancement of the memory will become very important and lensing might also need to be included. The gravitational wave community is currently studying the possibility of upgrading the current aLIGO detectors (e.g.~Voyager, Cosmic Explorer, Einstein Telescope) within the next one or two decades~\cite{Hild:2010id,Punturo:2010zz}. Such detectors will have a significantly improved sensitivity that will allow for the detection of gravitational waves emitted at much larger redshift. For such events, the redshift magnification will enhance the gravitational wave amplitude by an order of magnitude, while lensing may affect it by ${\cal{O}}(10\%)$. Recent work that included the redshift magnification suggests that such third-generation detectors may be able to detect the memory without need of stacking (ie.~with single events)~\cite{berti-favata}. The statistical combination of multiple events detected with third-generation observatories may also allow for the mapping of the lensing potential, although it is not clear how important this modification in the memory part of the signal will be. 

Space-borne detectors, such as LISA~\cite{Audley:2017drz}, are also being planned by both the European Space Agency and NASA, with an expected launch date of early 2030s. Such space-borne detectors will detect gravitational waves in the milliHz frequency range, allowing for the detection of supermassive black hole mergers (with masses as large as ${\cal{O}}(10^{7} M_{\odot}$) at redshifts as large as $10$. Clearly, for such events the redshift magnification will increase the amplitude of the signal by an order of magnitude and the focusing or defocusing effect of lensing will also be important. Recent work has suggested that such observations will also be able to detect the memory effect with single events~\cite{berti-favata}. Here again, the inclusion of the redshift magnification to the memory is crucial, although it is less clear how relevant the lensing correction is. Fortunately, the LISA community already has waveform models that include both the redshift magnification and the lensing correction calculated in this paper, so no additional model-building is necessary. It would be interesting to see if the lensing correction to the memory can contribute to the mapping of the lensing potential with many LISA observations, as has been argued could be possible with the non-memory part of the signal~\cite{Laguna:2009re,Holz:2005df}.

Last but not least, gravitational waves may also soon be detected in the nano-Hz range with pulsar timing arrays~\cite{Arzoumanian:2015liz}. The idea here is to cross-correlate the signal of multiple pulsars to disentangle any correlated residuals in the times of arrival of the pulses. Because of their frequency of operation, pulsar timing arrays are expected to detect gravitational waves produced in the mergers of supermassive black holes (with masses of ${\cal{O}}(10^{9}--10^{10})$ at redshifts of ${\cal{O}}(5--10)$. Once again, the inclusion of the redshift enhancement and the lensing correction to the memory should also be important in the extraction of the memory from pulsar-timing gravitational wave observations. Work along this line has only recently begun and is currently ongoing.   

\section*{Acknowledgements}
We would like to thank Bob Wald and Abhay Ashtekar for helpful discussions.  LB is supported by NSF grant DMS-1253149 to The University of Michigan. DG is supported by NSF grant PHY-1505565 to Oakland University. NY is supported by NSF CAREER Grant PHY-1250636 and NASA grant NNX16AB98G to Montana State University.

\appendix
\section{Propagation of Gravitational Waves on an FRLW Background}


For a compact binary inspiral, the wave zone, GW metric perturbation can be projected to a $+$ and $\times$ basis to obtain (to leading post-Newtonian order)
\begin{align}
\label{eq:hp}
h_{+}^{\Mink}(t)  &= \left[\frac{{\cal{M}}^{5/3}}{r_{\Mink}} f_{s}(t)^{2/3}\right] \cos\left[2 i \Phi(t)\right] e_{+}\,,
\\
\label{eq:hc}
h_{\times}^{\Mink}(t)  &= \left[\frac{{\cal{M}}^{5/3}}{r_{\Mink}} f_{s}(t)^{2/3}\right] \sin\left[2 i \Phi(t)\right] e_{\times}\,, 
\end{align}
where $e_{+,\times}$ are projections of the polarization basis tensors on the detector's antenna pattern tensor. The term in square brackets is the \emph{chirping amplitude}, which depends on the time-dependent frequency at the source $f_{s}(t)$, the flat-space coordinate distance from the source to the observer $r_{\Mink}$ and the chirp mass ${\cal{M}} = \eta^{3/5} m$, with $\eta = m_{1} m_{2}/m^{2}$ the symmetric mass ratio and $m = m_{1} + m_{2}$ the total mass. The oscillatory part of the GW is a function of the time-dependent orbital phase $\Phi(t)$, which will not play an important role in this discussion.

The GW modes presented above arise from solving the linearized Einstein equations in a given (Lorenz) gauge through Green functions with asymptotically flat boundary conditions, i.e.~a \emph{no-incoming wave condition} that requires the metric asymptotes to Minkowski spacetime $\eta_{ab}$ at future null infinity. Earth, however, is not in the wave zone where one can approximate spacetime as Minkowski, but rather it is in the cosmological zone, where spacetime is described by an FLRW metric. The line element of the latter can be written in the form
\be
\label{eq:ds2-FRW}
ds^{2}_{\FLRW} = a^{2}(\eta) \eta_{ab} dx^{a} dx^{b}\,,
\ee
where $\eta$ is conformal time. 

How does one then modify the GW solutions of Equations~(\ref{eq:hp}-\ref{eq:hc}) to account for their propagation on an FLRW background? The simplest way to do it is to use a recipe attributed to Thorne~\cite{Deruelle:1984hq}, which actually derives from an analysis of the evolution of the metric perturbation in an FLRW spacetime. The recipe consists of performing the following actions on 
Equations~(\ref{eq:hp}-\ref{eq:hc}): 
\begin{enumerate}
\item Replace $r_{\Mink}$ with the luminosity distance $D_{L}$,
\item Replace $f_{s}$ with the observed frequency $f_{o}$,
\item Replace ${\cal{M}}$ with the redshifted chirp mass ${\cal{M}}_{z} = (1 + z) {\cal{M}}$,
\end{enumerate}
which then leads to 
\begin{align}
\label{eq:hp-Thorne}
h_{+}^{\FLRW}(t)  &= \left[\frac{{\cal{M}}_{z}^{5/3}}{d_{L}} f_{o}(t)^{2/3}\right] \cos\left[2 i \Phi(t)\right] e_{+}\,,
\\
\label{eq:hc-Thorne}
h_{\times}^{\FLRW}(t)  &= \left[\frac{{\cal{M}}_{z}^{5/3}}{d_{L}} f_{o}(t)^{2/3}\right] \sin\left[2 i \Phi(t)\right] e_{\times}\,, 
\end{align}

To our knowledge, a mathematical explanation of this recipe dates back to~\cite{Deruelle:1984hq}, although different parts of the analysis have been rediscovered over the years~\cite{Alexander:2007kv,Tolish:2016ggo}. Less mathematical explanations have certainly appeared in the literature. From a physical viewpoint, the Minkowski radial distance $r_{\Mink}$ is nothing but the co-moving proper distance $d_{M}$, and when converting this to the luminosity distance via $d_{M} = d_{L} (1 + z)^{-1}$ and the observed frequency to the source frequency via $f_{s} = (1 + z) f_{o}$, one obtains Eqs.~\eqref{eq:hp-Thorne}-\eqref{eq:hc-Thorne}. From a field-theory viewpoint, the GWs we observe in the cosmological zone are given by the action of a plane-wave propagator acting on the GWs in the wave zone, namely
\be
h_{+,\times}^{\FLRW}(t) = \hat{P} \left[ h_{+,\times}^{\Mink}(t) \right] = \frac{e^{-i \omega t}}{1 + z} h_{+,\times}^{\Mink}(0)\,,
\ee
where $\omega$ is the angular frequency of the GW and the factor of $1+z$ arises due due to Hubble dilution. 

In what follows we put several pieces that have appeared in the literature together to provide a modern, pedagogical and mathematical derivation of Thorne's recipe. We begin by considering the propagation of a GW in a FLRW universe. Let the line element be
\be
\label{eq:ds2-FRW+wave}
ds^{2}= a^{2}(\eta) \left(\eta_{ab} + h_{ab}^{\FLRW} \right) dx^{a} dx^{b}\,,
\ee
and insert this metric into the Einstein equations, expanding in $|h_{ab}| \ll |\eta_{ab}|$.  We will now assume that the fluid equations decouple from the metric perturbation equations in the high-frequency limit, as we demonstrated in Sec.~\ref{sec:short-waves}. With this assumption in mind and to zeroth-order in perturbation theory, one finds the Friedman equations for the scale factor, which we can solve for any energy component for the Universe. To first order in perturbation theory, one finds
\be
\label{eq:wave-prop}
\square_{\FLRW} h_{ab}^{\FLRW} = 0\,,
\ee
after imposing the Lorenz gauge. The differential operator on FLRW is simply
\be
\square_{\FLRW} = \partial_{\eta}^{2} - \delta^{ij} \partial_{ij} + 2 {\cal{H}} \partial_{\eta}\,,
\ee
where ${\cal{H}} = a'/a$ and primes denote partial differentiation with respect to conformal time. 

Let us now use the geometric optics approximation (also known as the short-wavelength approximation or the WKB approximation) to express an ansatz for the solution to this differential equation:
\be
\label{eq:wave-ansatz}
h_{ab}^{\FLRW} = A_{ab}(\eta) e^{-i [\phi(\eta) - \kappa n_{k} \chi^{k}]}\,,
\ee
where $\kappa$ is the conformal wavenumber, $n_{k}$ is a spatial unit vector (pointing in the direction of propagation of the wave), $\chi^{i}$ is a conformal spatial coordinate and $\phi(\eta)$ is assumed to vary much more rapidly than the amplitude tensor $A_{ab}(\eta)$, i.e.~$A'/A \ll \phi'$. With this ansatz, the propagation equation reduces to
\be
\left[\kappa^{2} - 2 i {\cal{H}} \phi' - (\phi')^{2} - i \phi''\right] h_{ab} = 0\,.
\ee
Alternatively, one can also arrive at this equation through a Fourier analysis of Eq.~\eqref{eq:wave-prop}. 

The dispersion relation in Eq.~\eqref{eq:wave-prop} is a differential equation for $\phi$. One can solve this equation easily if one assumes that $\phi'' \ll (\phi')^{2}$ and that ${\cal{H}}' \ll {\cal{H}}^{2}$. The second condition is true in cosmology, unless one is considering the inflationary era. The first condition is true when $f' \ll f^{2}$; this, for example, holds during the inspiral of compact binaries. With these conditions, one then finds that
\be
\phi' = - i {\cal{H}} \pm \kappa  \left(1 - \frac{{\cal{H}}^{2}}{\kappa^{2}}  \right)^{1/2} = \pm \kappa - i {\cal{H}} + {\cal{O}}\left(\frac{{\cal{H}}^{2}}{\kappa}\right)\,, 
\ee
since ${\cal{H}}/\kappa \ll 1$ for the sources we have in mind, and then 
\begin{align}
\phi &= \pm \kappa \eta - i \int_{s}^{o} \; {\cal{H}} d \eta   
= \pm \kappa \eta - i  \ln \left(\frac{a_{o}}{a_{s}}\right)  \,.
\end{align}
for $\kappa$ nearly constant. 

With this at hand, we can now reconstruct the propagated GW: 
\begin{align}
h_{ab}^{\FLRW} &= A_{ab}(\eta) \; e^{-i  \kappa (\pm \eta  - n_{k} \chi^{k}) - \ln \left(\frac{a_{o}}{a_{s}}\right)}\,,
\nonumber \\
&= \frac{A_{ab}(\eta)}{1 + z} e^{-i  \kappa (\pm \eta  - n_{k} \chi^{k})}\,,
\end{align}
where we have used that ${a_{s}}/{a_{o}} = (1 + z)^{-1}$. This result is sensible because one expects the GW amplitude to be Hubble diluted as the GW propagates in an expanding background.

Let us then reconstruct the full GW, including the source dependence, as observed a cosmological distance away. Comparing Eq.~\eqref{eq:wave-ansatz} to Eqs.~\eqref{eq:hp} and~\eqref{eq:hc}, we see that $A_{ab} = {\cal{A}} \; e_{ab}$, where $e_{ab}$ is a polarization tensor and 
\be
{\cal{A}} = \frac{{\cal{M}}^{5/3}}{r_{\FLRW}} f_{s}^{2/3}\,. 
\ee
The radial distance here, $r_{\FLRW}$, is that associated with the FLRW conformal coordinate in Eq.~\eqref{eq:ds2-FRW+wave}, and thus, $r_{\FLRW} = d_{M} = (1 + z)^{-1} d_{L}$, where $d_{M}$ is the comoving distance and $d_{L}$ is the luminosity distance. We then have
\be
{\cal{A}} = \frac{{\cal{M}}^{5/3}  (1 + z)}{d_{L}} f_{s}(t)^{2/3}\,,
\ee
and re-expressing this result in terms of the observable frequency, $f_{s} = (1 + z) f_{o}$, one then finds
\begin{align}
{\cal{A}} &= \frac{{\cal{M}}^{5/3}}{d_{L}} (1 + z)^{5/3} f_{o}(t)^{2/3} = \frac{{\cal{M}}_{z}^{5/3}}{d_{L}} f_{o}(t)^{2/3}
\end{align}
One then sees that this is identical to what one finds with Thorne's trick, namely Eqs.~\eqref{eq:hp-Thorne} and~\eqref{eq:hc-Thorne}. 

\section{Amplitude of gravitational waves in an inhomogeneous universe}

We want to find the amplitude $\cal A$ satisfying
\be
{k^a}{\nabla _a}{\cal A} = -(1/2) {\cal A } \theta
\label{amplitude}
\ee
in the perturbed FLRW metric of eqn. (\ref{gFLRW-perturbed}).  Here $k^a$ is an affinely parameterized  null geodesic congruence with affine parameter $\lambda$,
and $\theta = {\nabla _a}{k^a}$ is the divergence of that congruence.  The metric of eqn. (\ref{gFLRW-perturbed}) can be written as
${a^2}( {\eta _{ab}}+{\gamma _{ab}})$ where $\eta_{ab}$ is the metric of Minkowski spacetime and 
\be
{\gamma_{ab}} = -2 [ (\Psi + \Phi) {t_a}{t_b} + \Psi {\eta _{ab}}] \; \; \; ,
\label{deltag}
\ee
and $t^a$ is the unit vector in the time direction.  

It is a standard result that if $k^a$ is an affinely parameterized null geodesic in the metric $g_{ab}$ then ${\Omega ^{-2}}{k^a}$ is an affinely parametrized null geodesic of the metric ${\Omega^2}{g_{ab}}$.  It then follows that if $\cal A$ is a solution of eqn. 
(\ref{amplitude}) for the metric $g_{ab}$ then ${\Omega^{-1}}{\cal A}$ is a solution of eqn. (\ref{amplitude}) in the metric 
${\Omega^2}{g_{ab}}$.  In this appendix we will calculate $\cal A$ for the perturbed flat metric ${\eta _{ab}}+{\gamma _{ab}}$
(see eqns. (\ref{amplituderesult}-\ref{xiresult}) below).  It then follows that $\cal A$ for the perturbed FLRW metric of eqn. (\ref{gFLRW-perturbed})
is the result of eqn. (\ref{amplituderesult}) multiplied by $a^{-1}$.

We start by calculating $\theta$ in the perturbed metric.  We do this two ways: first a coordinate method, like that of \cite{Laguna:2009re}, then a geometric method. 

We decompose $k^a$ as
\be
{k^a} = (1+\beta){{\bar k}^a} + \alpha {t^a} + {s^a}
\ee
where ${{\bar k}^a}={t^a}+{{\hat r}^a}$ is the background value of $k^a$, and $s^a$ is orthogonal to both $t^a$ and $k^a$.  We will calculate only to first order in perturbation theory.  Note that $\alpha, \, \beta$ and $s^a$ are all first order quantities.  Therefore to zeroth order in perturbation theory, we can use $k^a$ and ${\bar k}^a$ interchangably.  Also note that ${{\bar k}^a}{\nabla _a}r=1$.  Therefore to zeroth order in perturbation theory we can use $r$ and the affine parameter $\lambda$ interchangably.  

The fact that 
$g_{ab}{k^a}{k^b}=0$ immediately yields
\be
\alpha = - (\Phi + \Psi)
\ee
A decomposition of the Christoffel symbols yields
\bea
{\Gamma ^c _{ab}}{{\bar k}^a}{{\bar k}^b} &=& {t^c}{\partial _\lambda}(\Phi + \Psi) + {{\bar k}^c}({\partial _\lambda}(\Phi - \Psi) - {\partial _t} (\Phi + \Psi)) 
\nonumber
\\
&+& {P^{ca}}{\partial _a}(\Phi + \Psi)
\eea
where ${P^{ab}}={\delta ^{ab}}-{{\hat r}^a}{{\hat r}^b}$ is the projection to the two-sphere.
The geodesic equation then yields 
\bea
{\partial _\lambda} \alpha = - {\partial _\lambda}(\Phi + \Psi)
\nonumber
\\
{\partial _\lambda } \beta = {\partial _\lambda} (\Psi - \Phi) + {\partial _t} (\Phi + \Psi)
\nonumber
\\
{{\cal L}_k}{s^a} = - 2 {\lambda ^{-1}}{s^a} - {P^{ab}}{\partial _b}(\Phi + \Psi)
\eea
We then find
\bea
{\nabla _a}{k^a} &=& {\partial _a}{k^a} + {\Gamma ^a _{ab}}{k^b} 
\nonumber
\\
&=& {\partial _a}\left ( (1+\beta){{\bar k}^a} + \alpha {t^a} + {s^a} \right )  + {\Gamma ^a _{ab}}{{\bar k}^b}
\nonumber
\\
&=& {\frac 2 r}(1+\beta) + {\partial _\lambda}\beta + {\partial _t}\alpha + {\partial _a}{s^a} + {\partial _\lambda }(\Phi - 3 \Psi)
\nonumber
\\
&=& {\frac 2 r}(1+\beta) - 2 {\partial _\lambda} \Psi + {\partial _a}{s^a}
\label{divk}
\eea
However, we have
\bea
{{\cal L}_k}({\partial _a}{s^a}) &=& {\partial _a} \left ( {{\cal L}_k}{s^a} \right ) 
\nonumber
\\
&=&  {\partial _a} \left ( - 2 {\lambda ^{-1}}{s^a} - {P^{ab}}{\partial _b}(\Phi + \Psi) \right ) 
\nonumber
\\
&=& - 2 {\lambda ^{-1}} {\partial _a}{s^a} - {\lambda ^{-2}} {D_A}{D^A}(\Phi + \Psi)
\eea
where ${D_A}{D^A}$ is the Laplacian on the unit two-sphere.
We then find
\be
{\partial _a}{s^a} = - {\lambda ^{-2}} {\int _0 ^\lambda } \;  d {\lambda '} \; {D_A}{D^A}(\Phi + \Psi)   \; \; \; 
\label{divs}
\ee
We also have 
\be
{\frac {dr} {d \lambda}} = {k^a}{\nabla _a} r = 1 + \beta
\label{drdlambda}
\ee
Using eqns. (\ref{divs}) and (\ref{drdlambda}) in eqn. (\ref{divk}) we obtain
\be
\theta = {\frac 2 r} \; {\frac {dr} {d\lambda}} - 2 {\partial _\lambda}\Psi - {\lambda ^{-2}} {\int _0 ^\lambda } \;  d {\lambda '} \; {D_A}{D^A}(\Phi + \Psi)   \; \; \; 
\label{thetaresult}
\ee
Integrating eqn. (\ref{amplitude}) we then obtain
\be
{\cal A} = {\frac c r} \;  ( 1 + \zeta ) \; \; \; ,
\label{amplituderesult}
\ee
where the integration ``constant'' $c$ can depend on the angle, and the quantity $\zeta$ is given by
\bea
\zeta &=& \Psi - {\Psi _e} 
\nonumber
\\
&+& {\frac 1 2} {\int _0 ^\lambda } \; {\frac {d {\lambda '}} {{({\lambda '})}^2}} \; 
{\int _0 ^{\lambda '} } \;  d {\lambda ''} \; {D_A}{D^A}(\Phi + \Psi)  \; \; \; .
\label{xiresult}
\eea

We now turn to a geometric derivation of the result for $\cal A$.
The null Raychaudhuri equation is
\be
{\frac {d\theta} {d\lambda}} = - {\frac 1 2} {\theta ^2} - {\sigma _{ab}}{\sigma ^{ab}} + {\omega _{ab}}{\omega ^{ab}} - {R_{ab}}{k^a}{k^b}
\ee
But our null geodesic congruence is the light cone of the point of emission of the waves.  So the shear $\sigma _{ab}$ and the rotation 
$\omega _{ab}$ vanish in the flat spacetime background, and their squares are second order and can therefore be neglected in first order perturbation theory.  Thus to first order we have
\be
{\frac {d\theta} {d\lambda}} = - {\frac 1 2} {\theta ^2} - {R_{ab}}{k^a}{k^b}
\ee
which can be written as
\be
{\frac d {d \lambda}} ( {{\theta}^{-1}} ) = {\frac 1  2} + {\theta ^{-2}} {R_{ab}}{k^a}{k^b}
\ee
Thus we find
\bea
{\theta ^{-1}} &=& {\frac \lambda 2} + {\frac 1 4}  {\int _0 ^\lambda} {{({\lambda '})}^2} {R_{ab}}{k^a}{k^b} \; d {\lambda '}
\nonumber
\\
&=& {\frac \lambda 2} \left ( 1 + {\frac 1 {2\lambda }}  {\int _0 ^\lambda} {{({\lambda '})}^2} {R_{ab}}{k^a}{k^b} \; d {\lambda '} \right )
\eea
We therefore have
\bea
\theta &=& {\frac 2 \lambda} {{\left ( 1 + {\frac 1 {2\lambda }}  {\int _0 ^\lambda} {{({\lambda '})}^2} {R_{ab}}{k^a}{k^b} \; d {\lambda '} \right ) }^{-1}}
\nonumber
\\
&=& {\frac 2 \lambda} \; - \; {\frac 1 {\lambda ^2}}  {\int _0 ^\lambda} {{({\lambda '})}^2} {R_{ab}}{k^a}{k^b} \; d {\lambda '}
\eea
However we have 
\bea
&&{\frac 2 \lambda} \; - \; {\frac 2 r} \; {\frac {dr} {d\lambda}} = {\frac 2 \lambda} \; - \; {\frac 2 r} \; (1+\beta) 
\nonumber
\\
&=& {\frac 2 {\lambda r}} (r - \lambda ) - {\frac {2\beta} \lambda} 
\nonumber
\\
&=& - {\frac {2\beta} \lambda} \; + \; {\frac 2 {\lambda ^2}} {\int _0 ^\lambda} 
\; \left ( {\frac {dr} {d {\lambda '}}} - 1 \right ) \; d {\lambda '}
\nonumber
\\
&=& - {\frac {2\beta} \lambda} \; + \; {\frac 2 {\lambda ^2}} {\int _0 ^\lambda} \; \beta \; d {\lambda '}
\nonumber
\\
&=& - \; {\frac 2 {\lambda ^2}} {\int _0 ^\lambda} \; {\lambda '} \; {\frac {d\beta} {d {\lambda '}}} \; d {\lambda '}
\nonumber
\\
&=& - \; {\frac 2 {\lambda ^2}} {\int _0 ^\lambda} \; {\lambda '} \; \left ( {\partial _{\lambda '}} (\Psi - \Phi) + {\partial _t} (\Phi + \Psi) \right ) \; d {\lambda '}
\eea
We therefore find
\bea
\theta \; &-& \; {\frac 2 r} \; {\frac {dr} {d\lambda}} = - \; {\frac 1 {\lambda ^2}}  {\int _0 ^\lambda} \;  {{({\lambda '})}^2} \biggl ( {R_{ab}}{k^a}{k^b} 
\nonumber
\\
&+& {\frac 2 {\lambda '}} \left ( {\partial _{\lambda '}} (\Psi - \Phi) + {\partial _t} (\Phi + \Psi) \right )  \biggr ) 
\; d {\lambda '}
\label{thetaprelim}
\eea

A standard formula for the perturbed Ricci tensor is
\be
{R_{ab}} = {\partial ^c}{\partial _{(b}}{\gamma _{a)c}} \; - \; {1 \over 2} {\partial ^c}{\partial _c}{\gamma _{ab}} \; - \; 
{1 \over 2} {\partial _a}{\partial _b} \gamma \; \; \; .
\label{deltaRic}
\ee
Applying eqn. (\ref{deltaRic}) to the metric perturbation in eqn. (\ref{deltag}) yields
\bea
{R_{ab}} &=& {t_a}{t_b} {\partial ^c}{\partial _c}(\Psi + \Phi) \; + \; {\eta _{ab}} {\partial ^c}{\partial _c} \Psi
\; + \; {\partial _a}{\partial _b} (\Psi - \Phi) 
\nonumber
\\
&-& 2 {t_{(a}}{\partial _{b)}}{\partial _t}(\Psi + \Phi) \; \; \; .
\eea
Thus we find
\bea
{R_{ab}}{k^a}{k^b} &=& {\partial ^c}{\partial _c}(\Psi + \Phi) \; + \; {\frac {d^2} {d {\lambda ^2}}} (\Psi - \Phi) 
\nonumber
\\
&+& \;
2 {\frac d {d \lambda}} {\partial _t} (\Psi + \Phi) 
\nonumber
\\
&=&  2 {\frac {{d^2} \Psi} {d {\lambda ^2}}} \; + \; {\frac 2 \lambda} ( {\partial _\lambda}(\Psi+\Phi) - {\partial _t} (\Psi + \Phi)) 
\nonumber
\\
&+& {\lambda ^{-2}}{D_A}{D^A}(\Psi+\Phi) \; \; \; .
\eea
We then find 
\bea
{R_{ab}}{k^a}{k^b} &+& {\frac 2 \lambda} \left ( {\partial _\lambda} (\Psi - \Phi) + {\partial _t} (\Phi + \Psi) \right ) 
\nonumber
\\
&=& {\lambda ^{-2}} \left [2 {\frac d {d\lambda}} ({\lambda ^2} {\partial _\lambda } \Psi) + {D_A}{D^A}(\Psi+\Phi) \right ] \; \; \; .
\label{Rickk}
\eea
Using eqn. (\ref{Rickk}) in eqn. (\ref{thetaprelim}) then yields the result of eqn. (\ref{thetaresult}).  Thus the geometric method agrees with the coordinate method.
 

\end{document}